\documentclass[twocolumn,aps, prl,showpacs]{revtex4}

\usepackage{graphics}
\usepackage{graphicx}
\usepackage{dcolumn}
\usepackage{bm}
\usepackage{amsmath,amssymb}

\newcommand{\de}{\partial}

\newcommand{\eq}[2]{\begin{equation} \label{#1} #2 \end{equation}}

\begin{document}

\title{Jackiw-Rebbi states in interfaced binary waveguide arrays with Kerr nonlinearity}
\author{Truong X. Tran, Hue M. Nguyen, and D\~{u}ng C. Duong}
\affiliation{Department of Physics, Le Quy Don Technical University, 236 Hoang Quoc Viet str., 10000 Hanoi, Vietnam}
\date{\today}

\begin{abstract}
We systematically investigate the optical analogs of quantum relativistic Jackiw-Rebbi states in binary waveguide arrays in the presence of Kerr nonlinearity with both self-focusing and self-defocusing cases. The localized profiles of these nonlinear Jackiw-Rebbi states can be calculated exactly by using the shooting method. We show that these nonlinear Jackiw-Rebbi states have a very interesting feature which is totally different from all other well-known nonlinear localized structures, including optical solitons. Namely, the profiles of nonlinear JR states with higher peak amplitudes can totally envelope the ones with lower peak amplitudes. We demonstrate that media with the positive nonlinear coefficient can support stable Jackiw-Rebbi states for a wide range of peak amplitudes, whereas media with the negative nonlinear coefficient are only able to support Jackiw-Rebbi states with low peak amplitudes. A general rule for the detuning of nonlinear Jackiw-Rebbi states in binary waveguide arrays is found.
\end{abstract}
\maketitle

\section{I. INTRODUCTION}
\label{Introduction}

Waveguide arrays (WAs) are unique systems possessing many fundamental discrete photonic phenomena, for instance, discrete diffraction \cite{nature,jones}, discrete solitons \cite{nature,christodoulides,kivshar}. Recently, it was shown that some important nonlinear phenomena usually associated to fiber optics, such as the emission of resonant radiation from solitons and soliton self-wavenumber shift can also occur in WAs \cite{tranresonant1,tranresonant2}, and the supercontinuum in both frequency and wave number domains can be generated in WAs as well \cite{tranresonant3}.

Waveguide arrays are also well-known for being intensively used in simulating fundamental effects in nonrelativistic quantum mechanics, for instance, photonic Bloch oscillations \cite{nature,pertsch,lenz} and Zener tunneling \cite{ghulinyan}. On the other hand, binary waveguide arrays (BWAs) - a special class of WAs consisting of two alternating different types of waveguides -  present a promising photonic system to investigate fundamental relativistic quantum mechanics phenomena emerging from the Dirac equation, e.g., {\em Zitterbewegung} \cite{zitterbewegung}, Klein paradox \cite{klein}, and Dirac solitons in the nonlinear regime in both one-dimensional \cite{trandirac1,trandirac2,trandirac4,trandirac5}, two-dimensional cases \cite{trandirac3}, and even in curved spacetime \cite{trandirac6}.

Recently, it has been shown in Ref. \cite{tranjr1} that at the interface of two BWAs having opposite signs of the so-called Dirac mass one can create the optical analogs of special states, well-known in the quantum field theory as {\em Jackiw-Rebbi} (JR) states \cite{jackiw}. The interaction between Dirac solitons and JR states in BWAs has been studied in Ref. \cite{tranjr2}. The JR states led to the prediction of the fundamental charge fractionalisation phenomenon which plays a central role in the fractional quantum Hall effect \cite{laughlin}. One of the most amazing features of the JR states is the topological nature of their zero-energy solution which has been interpreted as a precursor to topological insulators \cite{hasan}. Topological photonics has a great potential in the development of extremely robust optical circuits \cite{rechtsman}. Quite recently, the JR states in interfaced BWAs have been shown, as expected, to be also extremely robust under influence of strong disturbance such as the turning on/off of the nonlinearity, the linear transverse potential, and the oblique incidence \cite{tranjr3}. The photonic topological defect states on the edge of just a BWA were experimentally found in Ref. \cite{malkova}. The topological defect mode at the interface between two periodic dimer chains has also been investigated in Ref. \cite{blanco}. So far, some schemes have been proposed to realize the JR states, for instance, by using an atomic Fermi-Dirac gas loaded in a periodic optical lattice \cite{ruostekoski}, or by using so-called heavy solitons in a fermionic superfluid \cite{yefsah}. A photonic implementation of the JR model in a slow-light polaritonic setup has been proposed in Ref. \cite{angelakis}. The topological JR states on a dislocation in a two-dimensional photonic crystal have been investigated both theoretically and experimentally in Ref. \cite{li}. Quite recently, the photonic JR states in all-dielectric structures controlled by bianisotropy have also been observed both numerically and experimentally in Ref. \cite{gorlach}.

The above-mentioned JR states in interfaced BWAs analyzed in Ref. \cite{tranjr1} have been investigated in the \emph{linear regime} where their exact analytical solutions have been provided. However, the exact solutions for JRs in BWAs in the \emph{nonlinear} regime have not been found. Actually, in Ref. \cite{tranjr1} we have just shown the existence of nonlinear JR states by launching the linear solutions of JR states (multiplied by some arbitrary factors) into BWAs with Kerr nonlinearity, and then let the beam self-adjust its profiles during propagation. This way only can help us to prove the existence of nonlinear JR states at the end of the beam propagation in BWAs, but all the other important properties of nonlinear JR states in BWAs have been completely unexplored. In this paper we will use the so-called shooting method to calculate the exact profiles of JR states in BWAs in the regime of Kerr nonlinearity with both self-focusing and self-defocusing nonlinearity, and then systematically investigate their properties.

\section{II. GOVERNING EQUATIONS AND LINEAR SOLUTIONS OF JR STATES}
\label{linearcase}

In this Section, let us briefly introduce the governing equations in interfaced BWAs and the \emph{linear} exact solutions of JR states which have already been obtained in Ref. \cite{tranjr1}. Some results of these linear JR solutions will be necessary for further discussion of nonlinear JR states.

Light propagation in a discrete, periodic binary array of Kerr nonlinear waveguides can be described in the continuous wave regime by the following dimensionless coupled-mode equations (CMEs) \cite{sukhorukov,longhi1}:
\eq{CWCM}{i\frac{da_{n}}{dz}+\kappa[a_{n+1}+ a_{n-1}] - (-1)^{n} \sigma a_{n} +  \gamma |a_{n}|^{2}a_{n}=0,}
where $a_{n}$ is the electric field amplitude in the $n$th waveguide, $z$ is the longitudinal spatial coordinate, $2\sigma$ and $\kappa$ are the propagation mismatch and the coupling coefficient between two adjacent waveguides of the array, respectively, and $\gamma$ is the nonlinear coefficient of waveguides which is positive for self-focusing, but negative for self-defocusing media. In order to observe JR states one needs to use two BWAs which are placed adjacent to each other as illustratively shown in Fig. \ref{fig1}(a). We want to emphasize that if $n<0$ (for the left-hand side BWA), then $\sigma$ takes the constant value $\sigma_{1}$, whereas if $n\geq0$ (the right-hand side BWA) $\sigma$, then takes the constant value $\sigma_{2}$.

After setting $\Psi_{1}(n) = (-1)^{n}a_{2n}$ and $\Psi_{2}(n) = i(-1)^{n}a_{2n-1}$, and following the standard approach developed in Refs. \cite{zitterbewegung,longhi2} we can introduce the continuous transverse coordinate $\xi \leftrightarrow n$ and the  two-component spinor $\Psi(\xi,z)$ = $(\Psi_{1},\Psi_{2})^{T}$ which satisfies the 1D nonlinear Dirac equation \cite{trandirac1}:
\eq{diracequation}{i\de_{z}\Psi = -i\kappa\hat{\sigma}_{x}\de_{\xi}\Psi + \sigma\hat{\sigma}_{z}\Psi - \gamma G,} where the nonlinear terms $G \equiv (|\Psi_{1}|^{2}\Psi_{1},|\Psi_{2}|^{2}\Psi_{2})^{T}$;  $\hat{\sigma}_{x}$ and $\hat{\sigma}_{z}$ are the usual Pauli matrices. In quantum field theory the parameter $\sigma$ in the Dirac equation is often called the mass of the Dirac field (or Dirac mass), and this mass parameter can be both positive and negative.

In the linear case (i.e., when $\gamma$ = 0) if $\sigma_{1}<0$ and $\sigma_{2}>0$, the exact continuous JR solutions of Eq. (\ref{diracequation}) have already been derived in Ref. \cite{tranjr1} as follows:
\eq{solutioncontinuous}{ \Psi(\xi) = \sqrt{\frac{|\sigma_{1}\sigma_{2}|}{\kappa(|\sigma_{1}|+|\sigma_{2}|)}}\left(\begin{array}{cc} 1 \\ i \end{array}\right)e^{-|\sigma(\xi)\xi|/\kappa}.} Solution (\ref{solutioncontinuous}) is the exact one to  the continuous Eq. (\ref{diracequation}) in the linear case, but it is an approximate solution to the discrete Eqs. (\ref{CWCM}). Obviously, this approximation will become better if the beam width gets larger.

If $|\sigma_{1}|=|\sigma_{2}|=\sigma_{0}$, as shown in Ref. \cite{tranjr1}, one can easily get following exact localized solutions for the discrete Eqs. (\ref{CWCM}) without nonlinearity ($\gamma$ = 0) for the following two cases:

If $-\sigma_{1}=\sigma_{2}=\sigma_{0}>0$, one gets the following JR state of the 1st type \cite{tranjr1}:
\eq{solutiondiscrete1}{a_{n} = b_{n}e^{i\delta_{1}z},} where the detuning $\delta_{1} \equiv \kappa-\sqrt{\sigma_{0}^{2}+\kappa^{2}}$, $b_{n}$ is real and independent of the variable $z$, $b_{2n-1} = b_{2n}$. For $n\geq0$ one has the following relationship: $b_{2n}/b_{2n+1} = \alpha \equiv -[\sigma_{0}/\kappa + \sqrt{1+\sigma_{0}^{2}/\kappa^{2}}]$, whereas for $n<0$ one has: $b_{2n+1}/b_{2n} = \alpha$. Note that the central region (at waveguides $n$ = 0 and 1) for generating a JR state of the 1st type must have a positive value for $(-1)^{n}\sigma$ (see Fig. \ref{fig1}(b) at the central region for more details).

However, if $\sigma_{1}=-\sigma_{2}=\sigma_{0}>0$, one has the following JR state of the 2nd type \cite{tranjr1}:
\eq{solutiondiscrete2}{a_{n} = b_{n}e^{i\delta_{2}z},} where the detuning $\delta_{2} \equiv \kappa+\sqrt{\sigma_{0}^{2}+\kappa^{2}}$, $b_{n}$ is again real and independent of the variable $z$, $b_{2n-1} = b_{2n}$. For $n\geq0$ one has: $b_{2n}/b_{2n+1} = -\alpha$, whereas for $n<0$ one has: $b_{2n+1}/b_{2n} = -\alpha$. Note that the central region (at waveguides $n$ = 0 and 1) for generating a JR state of the 2nd type must have a negative value for $(-1)^{n}\sigma$.

\section{III. LOCALIZED JR STATES WITH KERR NONLINEARITY}
\label{cubic}

\begin{figure}[htb]
\centerline{\includegraphics[width=0.45\textwidth]{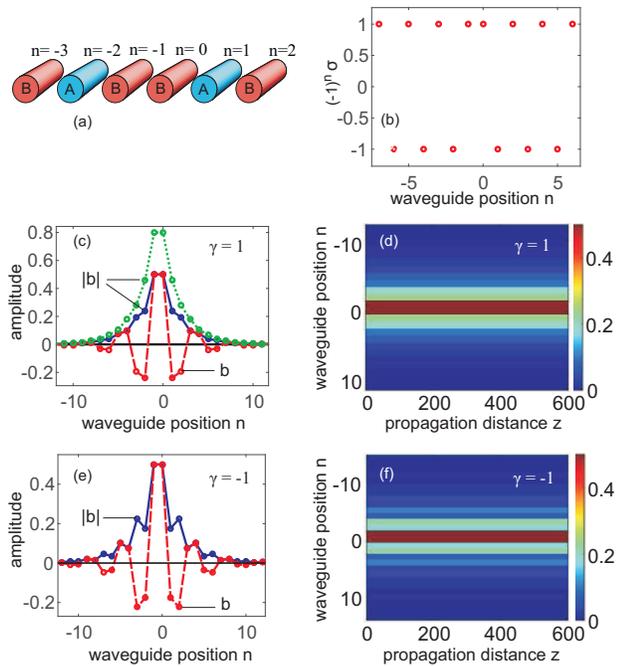}}
\caption{\small (color online). Localized JR states of the 1st type in interfaced BWAs with Kerr nonlinearity. (a) The scheme of two adjacent BWAs. (b) The value distribution of the array $(-1)^n\sigma$. (c) Amplitude profiles $b_{n}$ and $|b_{n}|$ when $\gamma = 1$. (d) Propagation of the nonlinear JR state in the (n,z)-plane with input condition taken from the dashed red curve in (c). (e) Amplitude profile $b_{n}$ and $|b_{n}|$ when $\gamma = -1$. (f) Propagation of the nonlinear JR state in the (n,z)-plane with input condition taken from (e). Parameters: $\sigma_{1}$ = -1; $\sigma_{2}$ = 1; $\kappa$ = 1.}\label{fig1}
\end{figure}

Now it is time for us to look for the nonlinear JR solutions in the presence of Kerr nonlinearity. Specifically, we will solve the nonlinear discrete Eqs. (\ref{CWCM}). In order to do that we will find the nonlinear JR solutions in the following form:
\eq{nonlinearsolution}{a_{n} = b_{n}e^{i\delta z},} where the amplitude $b_{n}$ is real and independent of the variable $z$ as in solutions (\ref{solutiondiscrete1}) and (\ref{solutiondiscrete2}). However, the detuning $\delta$ is the eigenvalue of each nonlinear JR state and must be found further. Of course, in the linear limit (i.e., when $\gamma$ = 0) the detuning $\delta$ will get the constant value of either $\delta_{1}$ or $\delta_{2}$ depending on whether one has the JR state of the 1st or 2nd type, respectively. After inserting the ansatz (\ref{nonlinearsolution}) into the coupled-mode equations (\ref{CWCM}) we will get the following system of algebraic equations:
\eq{algebraeq}{-\delta b_{n} = - \kappa[b_{n+1}+ b_{n-1}] + (-1)^{n} \sigma b_{n} -  \gamma |b_{n}|^{2}b_{n}.}

In Fig. \ref{fig1} we show the nonlinear JR state of the 1st type for the case of Kerr nonlinearity. The illustrative scheme of two BWAs with opposite propagation mismatches located adjacent to each other is presented in Fig. \ref{fig1}(a). The array $(-1)^n\sigma$ at the interface between the two BWAs is plotted in Fig. \ref{fig1}(b). Due to the symmetry of the system as shown in Figs. \ref{fig1}(a) and \ref{fig1}(b), we can find nonlinear solutions to Eqs. (\ref{algebraeq}) with the following property: $b_{n} = b_{-(n+1)}$, i.e., $b_{0} = b_{-1}$,  $b_{1} = b_{-2}$, and so on. With this property if the peak amplitude of the JR state at the central (i.e., $b_{0}$ and $b_{-1}$) is set, then all other values of $b_{n}$ can be calculated straightforwardly from Eqs. (\ref{algebraeq}). However, we are interested in finding the localized nonlinear JR solutions where the tails will vanish when $n \rightarrow \pm\infty$. In order to do that, we just need to refine the detuning $\delta$ until the conditions $b_{n} \rightarrow 0$ is held true when $|n|$ is large enough. Therefore, the eigenvalue of the detuning $\delta$ of the localized nonlinear JR states will be a function of their peak amplitude. With this simple and efficient "shooting" method, we are able to numerically find exact solutions of all kinds of JR states (including linear ones whose exact solutions have been found earlier as shown in Section 2) by shooting from the JR center to the tails. The similar shooting method has also been successfully exploited to numerically find the dissipative Bragg solitons in active nonlinear fibers in Ref. \cite{tranchaos} with the only big difference that the shooting therein has been conducted from one tail of the soliton to the other.

In Fig. \ref{fig1}(c) we plot the amplitude profiles of the localized JR states of the 1st type in the case of self-focusing nonlinearity ($\gamma = 1$). The dashed red curve and the solid blue curve in Fig. \ref{fig1}(c) represent, respectively, the amplitude profile $b_{n}$ and $|b_{n}|$ of a JR state with peak amplitude $b_{0}$ = $b_{-1} = 0.5$, and the eigenvalue of the detuning is found to be $\delta \simeq$ 0.5475 $\delta_{1}$. Meanwhile, the dotted green curve in Fig. \ref{fig1}(c) represents the amplitude profile of another JR state with peak amplitude $b_{0}$ = $b_{-1} = 0.8$, and the eigenvalue of the detuning is found to be $\delta \simeq$ -0.166 $\delta_{1}$. To verify these solutions we use them as initial conditions for solving Eqs. (\ref{CWCM}) along the propagation distance $z$. The propagation of the nonlinear localized JR state in the (n,z)-plane with input condition taken from the dashed red curve in Fig. \ref{fig1}(c) is demonstrated in Fig. \ref{fig1}(d). As clearly shown in Fig. \ref{fig1}(d), the profile of this nonlinear JR state is perfectly conserved during propagation. This JR state is perfectly stable, robust, and can propagate as long as we want without any distortion of its shape.

In Fig. \ref{fig1}(e) we show the amplitude profile $b_{n}$ and $|b_{n}|$ of the localized JR state of the 1st type in the case of self-defocusing nonlinearity ($\gamma = -1$) with peak amplitude $b_{0}$ = $b_{-1} = 0.5$, and the eigenvalue of the detuning is found to be $\delta \simeq$ 1.4537 $\delta_{1}$. The propagation of the nonlinear localized JR state in the (n,z)-plane with input condition taken from Fig. \ref{fig1}(e) is demonstrated in Fig. \ref{fig1}(f). In this case, the nonlinear JR state with self-defocusing nonlinearity is also perfectly stable. In order to obtain results in Fig. \ref{fig1} we use parameters as follows: $\sigma_{1}$ = -1; $\sigma_{2}$ = 1; $\kappa$ = 1; total number of waveguides $N$ = 61; the nonlinear coefficient $\gamma = 1$ in Figs. \ref{fig1}(c,d), but $\gamma = -1$ in Figs. \ref{fig1}(e,f).

To estimate real physical parameters of the calculated JR states below we use typical parameters in waveguide arrays made of AlGaAs \cite{morandotti2}, where the coupling coefficient and nonlinear coefficient in physical units are $K=1240m^{-1}$ and $\Gamma=6.5m^{-1}W^{-1}$, respectively. In this case, the power scale will be $P_{0}=K/\Gamma = 190.8 W$, thus the peak power of the JR state shown in Fig. \ref{fig1}(d) will be around 47$W$, and the length scale in the propagation direction will be $z_{0} = 1/K = 0.8 mm$.

\begin{figure}[htb]
\centerline{\includegraphics[width=0.45\textwidth]{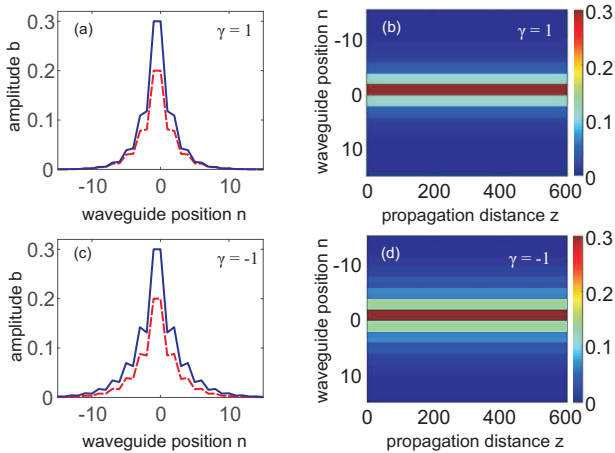}}
\caption{\small (color online). Localized JR states of the 2nd type in interfaced BWAs with Kerr nonlinearity. (a) Amplitude profile $b_{n}$ when $\gamma = 1$: the solid blue curve represents the JR state with peak amplitude $b_{0}$ = $b_{-1} = 0.3$, whereas the dashed red curve represents the JR state with peak amplitude $b_{0}$ = $b_{-1} = 0.2$. (b) Propagation of the nonlinear JR state in the (n,z)-plane with input condition taken from the solid blue curve in (a). (c) The same as (a), but for $\gamma = -1$. (d) Propagation of the nonlinear JR state in the (n,z)-plane with input condition taken from the solid blue curve in (c). Parameters: $\sigma_{1}$ = 1; $\sigma_{2}$ = -1; $\kappa$ = 1.}\label{fig2}
\end{figure}

In Figs. \ref{fig2}(a,b) we show the nonlinear localized JR states of the 2nd type with self-focusing nonlinearity, i.e., when $\gamma = 1$. The amplitude profiles $b_{n}$ of the JR states with $\gamma = 1$ in Fig. \ref{fig2}(a) are calculated by using the shooting method described above in solving Eqs. (\ref{algebraeq}) where the solid blue curve is the JR state with peak amplitude $b_{0}$ = $b_{-1} = 0.3$, and the eigenvalue detuning is found to be $\delta \simeq$  1.0285 $\delta_{2}$. Meanwhile, the dashed red curve in Fig. \ref{fig2}(a) is the JR state with peak amplitude $b_{0}$ = $b_{-1} = 0.2$, and the eigenvalue detuning is found to be $\delta \simeq$  1.0125$\delta_{2}$. The solid blue curve is used as initial conditions for solving Eqs. (\ref{CWCM}) and the evolution of $|a_{n}(z)|$ of this nonlinear localized JR state in the (n,z)-plane is demonstrated in Fig. \ref{fig2}(b).

Similarly, in Figs. \ref{fig2}(c,d) we show the nonlinear localized JR states of the 2nd type with self-defocusing nonlinearity, i.e., when $\gamma = -1$. Specifically, the solid blue curve in Fig. \ref{fig2}(c) is the amplitude profiles $b_{n}$ of the JR state with peak amplitude $b_{0}$ = $b_{-1} = 0.3$, and the eigenvalue detuning is found to be $\delta \simeq$ 0.9726 $\delta_{2}$; whereas the dashed red curve therein plots the JR state with peak amplitude $b_{0}$ = $b_{-1} = 0.2$, and the eigenvalue detuning is found to be $\delta \simeq$ 0.9877 $\delta_{2}$.  Figure \ref{fig2}(d) shows the propagation of the nonlinear localized JR state $|a_{n}(z)|$ in the (n,z)-plane with the input condition taken from the solid blue curve in Fig. \ref{fig2}(c). All parameters used for obtaining results in Fig. \ref{fig2} are as follows: $\sigma_{1}$ = 1; $\sigma_{2}$ = -1; $\kappa$ = 1; total number of waveguides $N$ = 61; the nonlinear coefficient $\gamma = 1$ in Figs. \ref{fig2}(a,b), but $\gamma = -1$ in Figs. \ref{fig2}(c,d).

Now we can see that one of common features of the nonlinear JR state profiles of both the 1st and 2nd types in the regime of Kerr nonlinearity with self-focusing ($\gamma$ = 1) is that $|b_{n}|$ (and as a result $|a_{n}|$) monotonically decreases from the JR center to each of its two tails as shown in both Fig. \ref{fig1}(c) and Fig. \ref{fig2}(a). To be more specific, in this case we have: $|b_{2n-1}| < |b_{2n}|$ if $n<0$, and $|b_{2n-1}| > |b_{2n}|$ if $n>0$. As a result, the nonlinear JR states profiles in Fig. \ref{fig1}(c) and Fig. \ref{fig2}(a) have just one peak at the center. On the contrary, for JR states of both the 1st and 2nd types in the regime of Kerr nonlinearity with self-defocusing ($\gamma$ = -1) the JR state profiles of $|b_{n}|$ are not monotonic functions from the JR center to each of its two tails. In this case, we have $|b_{2n-1}| > |b_{2n}|$ if $n<0$, and $|b_{2n-1}| < |b_{2n}|$ if $n>0$ (see both Fig. \ref{fig1}(e) and Fig. \ref{fig2}(c)). Therefore, the nonlinear JR states profiles in Fig. \ref{fig1}(e) and Fig. \ref{fig2}(c) have many peaks. As a result of this, if two nonlinear JR states of the same type have the same peak amplitude, then the JR state with self-focusing nonlinearity is more transversally localized than the JR state with self-defocusing nonlinearity. This tendency is especially more visible with JR states of the 2nd type. Indeed, Fig. \ref{fig2} clearly shows that, when comparing JR states with the same peak amplitude, the JR states with self-focusing nonlinearity demonstrated in Figs. \ref{fig2}(a,b) are more localized in the transverse direction than the JR states with self-defocusing nonlinearity presented in Figs. \ref{fig2}(c,d). This difference of nonlinear JR states between positive and negative $\gamma$ is important and will decrease the possibility of generating JR states with high peak amplitudes in the regime with negative $\gamma$. We will return to this point in Section 4.

It is worth mentioning that for linear JR states we always have $b_{2n-1} = b_{2n}$ (see Ref. \cite{tranjr1} for more details). As a result, the profile of the linear JR state shown in Fig. 1(c) in Ref. \cite{tranjr1} has many horizontal steps. The difference between $|b_{2n-1}|$ and $|b_{2n}|$ of nonlinear JR states will be greater and greater if we increase their peak amplitudes $|b_{0}|$, lesser and lesser if we decrease $|b_{0}|$. So, obviously, the profiles of nonlinear JR states will be more and more similar to the ones of linear JR states if we decrease their peak amplitudes $|{b_{0}}|$. This fact is also important and is related to the property of the nonlinear JR state detunings as will be discussed further.

From Fig. \ref{fig2} one can see that all components $b_{n}$ of the JR states of the 2nd type shown in Figs. \ref{fig2}(a,c) have the same phase, i.e., they are all positive. Meanwhile, for the JR states of the 1st type shown in Figs. \ref{fig1}(c,e) only $b_{2n}$ and $b_{2n-1}$ (for instance, $b_{0}$ and $b_{-1}$; $b_{2}$ and $b_{1}$) have the same phase, but $b_{2n}$ and $b_{2n+1}$ (for instance, $b_{0}$ and $b_{1}$; $b_{-2}$ and $b_{-1}$) have the opposite phases.

While analyzing Fig. \ref{fig1}(c) and Figs. \ref{fig2}(a,c) one can see a distinguishing feature of nonlinear JR state profiles (represented by $|b_{n}|$) which is totally different from all other well-known localized nonlinear structures such as solitons emerging from the nonlinear Schr\"{o}dinger equation (i.e., temporal solitons in a single optical fiber, spatial solitons in an optical waveguide \cite{agrawal}), Bragg solitons, discrete solitons in a conventional waveguide array (see Ref. \cite{kivshar} for more details), and even discrete Dirac solitons in a BWA found in Ref. \cite{trandirac1}. This distinguishing feature of nonlinear JR states in interfaced BWAs is the fact that the profiles with higher peak amplitudes can totally envelope the profiles with lower peak amplitudes. Indeed, in Fig. \ref{fig1}(c) the blue solid curve with a lower peak amplitude is completely enveloped by the dotted green curve with a higher peak amplitude. Similarly, in Figs. \ref{fig2}(a,c) the dashed red curves with lower peak amplitudes are totally enveloped by the solid blue curves with higher peak amplitudes. This situation cannot happen with all other well-known nonlinear localized structures just mentioned above, including optical solitons. This distinguishing feature of nonlinear JR states is universal for all kinds of JR states with Kerr nonlinearity investigated in this paper, including JR states of the 1st type with negative $\gamma$ in Fig. \ref{fig1}.

\section{IV. DETUNINGS OF LOCALIZED JR STATES WITH KERR NONLINEARITY}
\label{detunings}

\begin{figure}[htb]
\centerline{\includegraphics[width=0.45\textwidth]{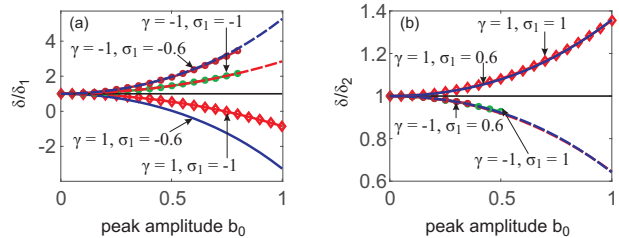}}
\caption{\small (color online). (a) Relative detuning $\delta/\delta_{1}$ of nonlinear JR states of the 1st type as a function of the peak amplitude $b_{0}$. (b) Relative detuning $\delta/\delta_{2}$ of nonlinear JR states of the 2nd type as a function of the peak amplitude $b_{0}$. All important parameters of each curve are indicated explicitly therein and also in the text.}\label{fig3}
\end{figure}

In this Section we will investigate the eigenvalue detuning $\delta$ of nonlinear JR states in detail. In Fig. \ref{fig3}(a) we plot the relative detuning $\delta/\delta_{1}$ of nonlinear JR states of the 1st type as a function of the peak amplitude $b_{0}$ for some sets of parameters. Specifically, the red curve with diamond markers is obtained when $\gamma$ = 1, $\sigma_{1}$ = -1, $\sigma_{2}$ = 1, whereas the green curve with round markers is obtained when we just change the sign of the nonlinear coefficient $\gamma$, i.e., when $\gamma$ = -1, $\sigma_{1}$ = -1, $\sigma_{2}$ = 1. These two curves are {\em almost} symmetrical with respect to the black horizontal axis representing the detuning $\delta$ = $\delta_{1}$ of the \emph{linear} JR states of the 1st type. In order to check the symmetry of these two curves, in Fig. \ref{fig3}(a) we also plot the dashed red curve which is the mirror image of the red curve with diamond markers with respect to the black horizontal axis. As clearly shown in Fig. \ref{fig3}(a), the dashed red curve and the green curve with round markers practically coincide with each other, especially for small values of $|b_{0}|$. The only big difference between them is the fact that the curve with self-focusing nonlinearity (the red one with diamond markers) can develop further to the right, whereas the curve with self-defocusing nonlinearity (the green one with round markers) stops at the maximum value of the peak amplitude $b_{0} \simeq 0.8$. This difference is related to the degree of transverse localization of nonlinear JR states discussed above. As already pointed out in Section 3, when we fix the peak amplitude and all other parameters of nonlinear JR states except for the sign of the nonlinear coefficient $\gamma$, then the profile of the nonlinear JR states with negative $\gamma$ will be less localized in the transverse direction than the one of the nonlinear JR states with positive $\gamma$. On the other hand, as also already mentioned above in Section 3, when the peak amplitude $|b_{0}|$ of JR states increases, their transverse dimension also gets larger. Note that here we do not use the term {\em beam width} because it is hard to define exactly this quantity for unsmooth profiles of JR states. As a result, when the peak amplitude $|b_{0}|$ increases, the nonlinear JR states will be less localized in the transverse direction, especially in the case of self-defocusing nonlinearity. This leads to the situation when $b_{0}$ is high enough ($b_{0} > 0.8$ in Fig. \ref{fig3}(a)), one still can obtain the localized solution for JR states with positive $\gamma$, but it is impossible to achieve it with negative $\gamma$.

In Fig. \ref{fig3}(a) we also show the dependence of the relative detuning $\delta/\delta_{1}$ of nonlinear JR states of the 1st type on the the peak amplitude $b_{0}$ for another value $\sigma_{1}$ = -0.6. The solid blue curve in Fig. \ref{fig3}(a) is the case when $\gamma$ = 1, $\sigma_{1}$ = -0.6, $\sigma_{2}$ = 0.6; and its mirror image with respect to the the black horizontal axis is the dashed blue curve. Meanwhile, the red curve with round markers is the case when we only change the sign of the nonlinear coefficient, i.e., when $\gamma$ = -1, $\sigma_{1}$ = -0.6, $\sigma_{2}$ = 0.6. We can see that all important properties of the relative detuning curve $\delta/\delta_{1}$ in the case when $\sigma_{1}$ = -1 analyzed above are reproduced again in the case when $\sigma_{1}$ = -0.6. Another feature of all curves plotted in Fig. \ref{fig3}(a) is the fact that they all begin from the origin at the point where $b_{0}$ = 0 and $\delta/\delta_{1}$ = 1. This is understandable because when the peak amplitude $b_{0}$ is small, nonlinear JR states will have properties which are more similar to the ones of linear JR states, and $\delta$ is closer to the detuning value $\delta_{1}$ of the linear JR states of the 1st type.

In Fig. \ref{fig3}(b) we plot the relative detuning $\delta/\delta_{2}$ of nonlinear JR states of the 2nd type as a function of the peak amplitude $b_{0}$ for some sets of parameters indicated therein. To be more specific, the red curve with diamond markers represents the case when $\gamma$ = 1, $\sigma_{1}$ = 1$, \sigma_{2}$ = -1, whereas the blue solid curve (which practically coincides with the red curve with diamond markers) is the case when $\gamma$ = 1, $\sigma_{1}$ = 0.6, $\sigma_{2}$ = -0.6. The mirror images of these two curves (with respect to the black horizontal line representing the detuning $\delta$ = $\delta_{2}$ of the linear JR states of the 2nd type) are also plotted as the dashed blue curve and the dashed red curve. Note that the dashed blue curve almost totally hides the dashed red curve, so the latter one can only be seen by enlarging Fig. \ref{fig3}(b) in the electronic version of this paper. The green curve with round markers and the red curve with round markers in Fig. \ref{fig3}(b) show the relative detuning curve in the case of self-defocusing nonlinearity when $\gamma$ = -1, $\sigma_{1}$ = 1, $\sigma_{2}$ = -1; and $\gamma$ = -1, $\sigma_{1}$ = 0.6, $\sigma_{2}$ = -0.6, respectively.

While comparing Fig. \ref{fig3}(a) with Fig. \ref{fig3}(b), one can see that all qualitative properties of the relative detuning curves discussed above in the case of JR states of the 1st type are reproduced again in the case of JR states of the 2nd type. However, the relative detuning curves for JR states of the 2nd type shown in Fig. \ref{fig3}(b) are practically independent of the value of $\sigma_{1}$ (as long as it is positive in order to obtain the JR states of the 2nd type), whereas they are more sensitive to the value of $\sigma_{1}$ in the case of JR states of the 1st type shown in Fig. \ref{fig3}(a) (which must be negative, of course, in order to obtain the JR states of the 1st type). Another difference is that with negative $\gamma$ one can only generate JR states of the 2nd type with rather low peak amplitudes $b_{0}$ in Figs. \ref{fig3}(b). Meanwhile, one can get JR states of the 1st type with higher peak amplitudes $b_{0}$ for negative $\gamma$ in Figs. \ref{fig3}(a). We suppose that this is because JR states of the 2nd type with negative $\gamma$ are much less localized in the transverse direction than all other types of nonlinear JR states. Therefore, it is more difficult to generate the localized nonlinear JR states of the 2nd type with high peak amplitudes when $\gamma$ = -1.

As mentioned above, the relative detuning of nonlinear JR states have an important feature: the two curves representing the relative detunings of nonlinear JR states for two opposite values $\gamma = 1$ and $\gamma = -1$ while fixing all other parameters are almost symmetrical with respect to the black horizontal axis in Fig. \ref{fig3} (which represents the detuning of the linear JR states). This important feature of nonlinear JR states can be proved quite straightforwardly in a general way as follows. Suppose that $\delta_{l}$ is the detuning parameter of {\em linear} JR states, i.e., $\delta_{l}$ = $\delta_{1}$ or $\delta_{2}$ for linear JR states of the 1st or 2nd type, respectively. Suppose also that $\delta$ is the detuning of the nonlinear JR states at the peak amplitude $b_{0}$ in the case of $\gamma$. Now we switch the sign of $\gamma$ so that $\gamma \rightarrow -\gamma$. By doing that the detuning of the nonlinear JR states will be transformed $\delta \rightarrow \delta^{'}$, and the amplitude will be changed $b_{n} \rightarrow c_{n}$. In order to have the situation when the curve representing $\delta/\delta_{l}$ is almost the mirror image of the curve representing $\delta^{'}/\delta_{l}$ with respect to the black horizontal line in Fig. \ref{fig3}, one must have the following relationship: $\delta/\delta_{l} \simeq 2 - \delta^{'}/\delta_{l}$. As a result, one has: $\delta^{'} \simeq 2\delta_{l} - \delta$. So, for the case of $-\gamma$, Eqs. (\ref{algebraeq}) will now have the following form:

\eq{algebraeq2}{-(2\delta_{l} - \delta)c_{n} \simeq - \kappa[c_{n+1}+ c_{n-1}] + (-1)^{n} \sigma c_{n} +  \gamma |c_{n}|^{2}c_{n}.} Now we add Eqs. (\ref{algebraeq}) to Eqs. (\ref{algebraeq2}) and get the following equations:

\eq{algebraeq3}{
\begin{split}
\delta (c_{n} - b_{n})& -2\delta_{l}c_{n} \simeq - \kappa[(c_{n+1}+b_{n+1}) + (c_{n-1} + b_{n-1})] + \\
&(-1)^{n} \sigma (c_{n} + b_{n}) +  \gamma (|c_{n}|^{2}c_{n} - |b_{n}|^{2}n_{n}).
\end{split}
} Now it is easy to see that if we fix the peak amplitudes so that $b_{0} = c_{0}$ and suppose that the condition $c_{n} \simeq b_{n}$ is held true (which is practically satisfied if the peak amplitudes are small, i.e., when we operate in the regime close to the linear case), then from Eqs. (\ref{algebraeq3}) we will have:

\eq{algebraeq4}{-\delta_{l}c_{n} \simeq - \kappa[c_{n+1}+ c_{n-1}] + (-1)^{n} \sigma c_{n}.} The latter equations are automatically satisfied in the regime close to the linear case, because parameter $\delta_{l}$ in Eqs. (\ref{algebraeq4}) is the detuning of the {\em linear} JR states as set above (see also Eqs. (\ref{algebraeq}) in the linear regime when $\gamma$ = 0). So, we have proved that  when changing the sign of the nonlinear coefficient $\gamma$ and fixing all other parameters one will obtain two curves representing relative detunings of nonlinear JR states which are almost symmetrical with respect to the horizontal axis representing the detuning of the \emph{linear} JR states. This tendency is held true very well when we operate in the regime close to the linear one, i.e., when the peak amplitude $|b_{0}|$ of the nonlinear JR states is small. On the contrary, this tendency will perform worst and worst if $|b_{0}|$ becomes larger, because in that case the condition $b_{n} \simeq c_{n}$ cannot be satisfied while fixing the peak amplitudes $b_{0} = c_{0}$. This important feature for the relative detuning of nonlinear JR states is universal. It is applicable not only for JR states in the regime of Kerr nonlinearity, but also for other types of nonlinearity, and even for other nonlinear states based on BWAs which will be published elsewhere.

\section{V. CONCLUSIONS}
\label{conclusions}

In this work we have systematically investigated the optical analogs of quantum relativistic Jackiw-Rebbi states in interfaced binary waveguide arrays in the regime of Kerr nonlinearity with both self-focusing and self-defocusing media. By using the shooting method we can numerically obtain the profiles of these nonlinear JR states which have a distinguishing feature that all other well-known localized nonlinear structures (including solitons) do not possess. Namely, if we increase nonlinear JR states peak amplitude, then their transverse dimension also increases at the same time. We have found out that JR states with self-focusing nonlinearity can exist in a wide range of their peak amplitudes, whereas the ones with self-defocusing nonlinearity can only exist in the low-peak amplitude regime. We have also systematically studied the detunings of these nonlinear JR states as functions of their peak amplitudes and demonstrated that by changing the sign of the nonlinear coefficient while fixing all other parameters one can obtain two curves for the relative detunings of nonlinear JR states which are almost symmetrical with respect to the axis representing the detuning of the corresponding linear JR states. This general rule is applicable not only for JR states with Kerr nonlinearity, but also for other types of nonlinearity, and even for other kinds of nonlinear localized states in binary waveguide arrays.

\section{ACKNOWLEDGMENT}
This research is funded by Vietnam National Foundation for Science and Technology Development (NAFOSTED) under grant number 103.03-2019.03.

\end{document}